\documentclass{jps-cp}
\usepackage{txfonts} 

\title{MR-ToF-assisted In-Source Resonance Ionization Spectroscopy at IGISOL} 

\author{Natalia \textsc{Ambrosio-Macias}$^{1}$, 
Daniel \textsc{Bettaney}$^{2}$ 
Saikumar \textsc{Chinthakayala}$^{1,4,5}$,  
Tobias	\textsc{Christen}$^{3}$, 
Marie \textsc{Deseyn}$^{3}$, 
Tommi \textsc{Eronen}$^{1}$, 
Zhuang \textsc{Ge}$^{1}$,
Phillip \textsc{Imgram}$^{3,6}$, 
Arthur \textsc{Jaries}$^{1, 7}$, 
Anu \textsc{Kankainen}$^{1}$, %
Brian  \textsc{Kootte}$^{1, 8}$, %
Markus \textsc{Kortelainen}$^{1}$, %
Sonja \textsc{Kujanpää}$^{1, 9}$, 
Kara \textsc{Lynch}$^{2}$,
Iain \textsc{Moore}$^{1}$,
Luis Miguel \textsc{Motilla}$^{4}$,
Andrea \textsc{Raggio}$^{1, 10}$,
Mikael \textsc{Reponen}$^{1}$,
Ville \textsc{Virtanen}$^{1, 11}$ 
and
Miikka \textsc{Winter}$^{1}$
}

\inst{{\footnotesize$^{1}$Department of Physics, Accelerator laboratory, University of Jyväskylä, P.O. Box 35(YFL), Jyväskylä, FI-40014, Finland\\
$^{2}$Schuster Building, University of Manchester, Brunswick Street, Manchester, M13 9PL, United Kingdom\\
$^{3}$KU Leuven, Instituut voor Kern- en Stralingsfysica, B-3001 Leuven, Belgium\\
$^{4}$GANIL, CEA/DRF-CNRS/IN2P3, Bd H. Becquerel, 14076 Caen, France\\
$^{5}$Université de Caen Normandie, Caen, France\\
$^{6}$Present Address: GSI Helmholtzzentrum f{\"u}r Schwerionenforschung GmbH, Darmstadt, 64291, Germany\\
$^{7}$Present Address: CERN, CH 1211 Geneva 23, Switzerland\\
$^{8}$Present Address: University of Northern British Columbia 3333 University Way Prince George, B.C.  Canada  V2N 4Z9 \\
$^{9}$Present Address: Laboratory for Biological Geochemistry, School of Architecture, Civil \& Environmental Engineering, École Polytechnique Fédérale de Lausanne, GR C2 505, Station 2, 1015 Lausanne, Switzerland\\
$^{10}$Present Address: KU Leuven, Instituut voor Kern- en Stralingsfysica, B-3001 Leuven, Belgium\\
$^{11}$Present Address: GANIL, CEA/DRF-CNRS/IN2P3, Bd H. Becquerel, 14076 Caen, France}}

\email{natalia.i.ambrosio{-}macias@jyu.fi}

\recdate{October 8, 2026}

\abst{Multi-Reflection Time-of-Flight Mass Spectrometer (MR-ToF-MS) assisted in-source laser resonance ionization spectroscopy (RIS) has been implemented at the Ion Guide Isotope Separator On-Line (IGISOL) facility, University of Jyväskylä. Exotic nuclei, produced via fusion-evaporation reactions, are stopped and neutralized in a hot-cavity catcher, and selectively laser-ionized through a multi-step resonance ionization scheme before being transported to the MR-ToF-MS device. This combination of techniques is aimed at optically probing the very exotic \textit{N=Z} nuclei in the immediate vicinity of $^{100}$Sn. The laser system, hot-cavity ion source, isobaric purification and detection stages are described. A representative hyperfine-structure spectrum of the proton-rich silver isotope $^{101}$Ag is presented as a case example.}

\kword{Resonance Ionization Spectroscopy, hyperfine structure, hot cavity ion source, MR-ToF.}

\begin{document}
\maketitle

\section{Introduction}

Laser spectroscopy of the atomic hyperfine structure yields hyperfine parameters and isotope shifts, which provide a nuclear-model-independent probe of ground- and isomeric-state properties, such as electromagnetic moments and changes in the mean-squared charge radii \cite{campbell2016, yang2023}. These observables offer a rigorous benchmark for current nuclear model approaches, for instance, $\it{ab\ initio}$ and density-functional-theory (DFT) methods, particularly in regions far from stability \cite{degroote2024, vernon2022, karthein2024}. 

Reaching the most exotic cases is however an experimental challenge: towards the $N$ = $Z$ line, production rates fall below one ion per second, half-lives decrease to hundreds or tens of milliseconds, and isobaric contamination from competing fusion-evaporation channels often overwhelms the ions of interest \cite{reponen2021, ge2024}. In-source resonance ionization spectroscopy (RIS), performed inside a heated ion-source cavity, combines the efficiency and selectivity needed to reach such cases, although at the cost of a spectroscopic resolution limited to several GHz by a combination of Doppler broadening and laser bandwidth \cite{fedosseev2012, reponen2015}. When combined with fast and precise mass filters, the effects of isobaric contamination can be mitigated \cite{reponen2021, eliseev2013}, enabling optical measurements with rates as low as an ion every few minutes.

In this contribution we present the implementation of MR-ToF-assisted RIS at IGISOL: production and laser ionization of isotopes in a hot-cavity catcher ion source, transport through the IGISOL beamline, and mass separation and detection using a newly commissioned multi-reflection time-of-flight mass spectrometer (MR-ToF-MS) \cite{virtanen2026mrtof}, together with representative hyperfine spectra for proton-rich silver isotopes. The hot cavity setup has already been utilized in probing the ground-state properties of exotic silver nuclei beyond the $N$ = 50 shell closure \cite{reponen2021, ge2024}. The nuclear-structure results derived from these spectra, including the magnetic dipole moments up to $^{95}$Ag, will be presented in a forthcoming publication.

\section{Experimental methodology}

The isotopes in the region of interest are usually produced via heavy-ion fusion-evaporation reactions, driven by primary beams from the K-130 cyclotron impinging on a thin-foil target mounted adjacent to the hot-cavity catcher laser ion source (HCLIS). The reaction products recoil out of the target and implant into a hot graphite catcher, enclosed in a  tantalum tube. The assembly is inductively heated to approximately 1800~K \cite{reponen2021, reponen2015}. The implanted ions diffuse out of the catcher as neutral atoms and effuse into a resistively heated carbon-based transfer tube, held at a similar temperature, which confines the atoms for efficient laser-atom interaction. The high operating temperature is essential for prompt diffusion of the recoiling atoms out of the graphite lattice, at the cost of a broadened atomic velocity distribution that ultimately limits the spectral resolution attainable with this technique.


\begin{figure}[tbh]
\centering
\includegraphics[width=1\linewidth]{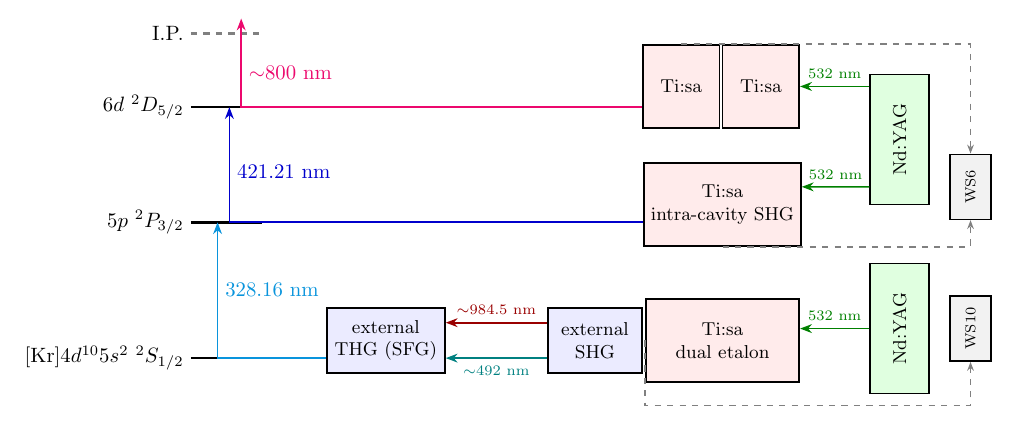}
\caption{Laser system employed for the three-step resonance ionization scheme of silver. The spectroscopy step, corresponding to the $4d\,{}^{10}5s\,{}^{2}S_{1/2} \rightarrow 5p\,{}^2P_{3/2}$ transition at 328.16~nm, is achieved with a dual-etalon cavity with external second-harmonic generation (SHG) and third-harmonic generation (THG) constructed by sum-frequency generation (SFG). A second cavity, with intra-cavity SHG, provides the second step at 421.21~nm. Two Ti:sa lasers, without wavelength selective elements, provide the third, non-resonant step at $\sim$800~nm with a combined power sufficient for saturation.}
\label{fig:laser}
\end{figure}

The atoms are then selectively laser-ionized through a multi-step resonance ionization process, for which Titanium:sapphire (Ti:sa) lasers are employed, pumped by 10~kHz Nd:YAG lasers. The FURIOS laser ion source \cite{reponen2013furios} has recently been upgraded to host two Photonics DM1-532-60 pump lasers to complement the existing Lee Laser LDP-200 MQG pump lasers. Moreover, the setup now hosts a DXG-532-40 short-pulse Nd:YAG laser used to drive non-resonant ionization transitions and a HighFinesse WS7 wavemeter in addition to the existing HighFinesse WS6 and WS10 wavemeters. A schematic layout of the laser system used for this work is shown in Fig.~\ref{fig:laser}.

The lasers are transported on individual paths to the front end of the IGISOL facility, where they are combined before a window on a vacuum chamber downstream from the extraction ion-optical elements. The lasers pass along the central axis of the IGISOL beamline, via extraction electrodes and a radiofrequency sextupole ion guide (SPIG) \cite{karvonen2008}, and are gradually focused into the HCLIS transfer tube where they overlap with the effusing atomic beam. For the resonant laser ionization of silver (Fig.~\ref{fig:laser}), the frequencies of the second and ionizing steps are monitored by a HighFinesse WS6 wavemeter, with the first step, used for the spectroscopy, monitored by a high-precision WS10 model.


Following laser ionization, the ions pass from the transfer tube into the SPIG, which transports them out of the source region to then be electrostatically accelerated to 30~keV and mass-separated by a dipole magnet, with a mass resolving power $\sim$ 400. The mass-selected continuous beam is injected into a helium gas-filled radiofrequency quadrupole (RFQ) cooler-buncher \cite{virtanen2025cooler}, which creates narrow ion bunches ($<$ 100 ns) according to a defined accumulation period, ready to be delivered to a multi-reflection time-of-flight mass spectrometer (MR-ToF-MS) \cite{virtanen2026mrtof}.

This recently commissioned mass-separation device provides mass-resolving powers of 1.5 $\times$ 10$^{5}$ within a 20~ms trapping time, and time-of-flight peak widths down to 22~ns FWHM. Compared with the JYFLTRAP Penning trap mass spectrometer used for phase-imaging ion-cyclotron-resonance (PI-ICR) assisted RIS\cite{reponen2021}, the MR-ToF-MS reaches a sufficient mass-resolving power on a millisecond timescale, rather than hundreds of ms. Statistics are accumulated with every single bunch of ions injected, highly advantageous for very exotic, short-lived cases, which would otherwise decay during longer Penning trap purification cycles. Ions of the mass of interest (in this case silver) are selected by a time gate on the MR-ToF-MS time-of-flight spectrum and counted as a function of the tagged spectroscopy-laser frequency. In this manner, the hyperfine spectrum can be obtained in an analogous way to that used in the PI-ICR approach wherein the software gating is performed on the ion landing position on the 2D MCP detector \cite{reponen2021, ge2024}.

\section{Results: hyperfine spectrum of $^{101}$Ag}

\begin{figure}[tbh]
\centering
\includegraphics[width=0.75\linewidth]{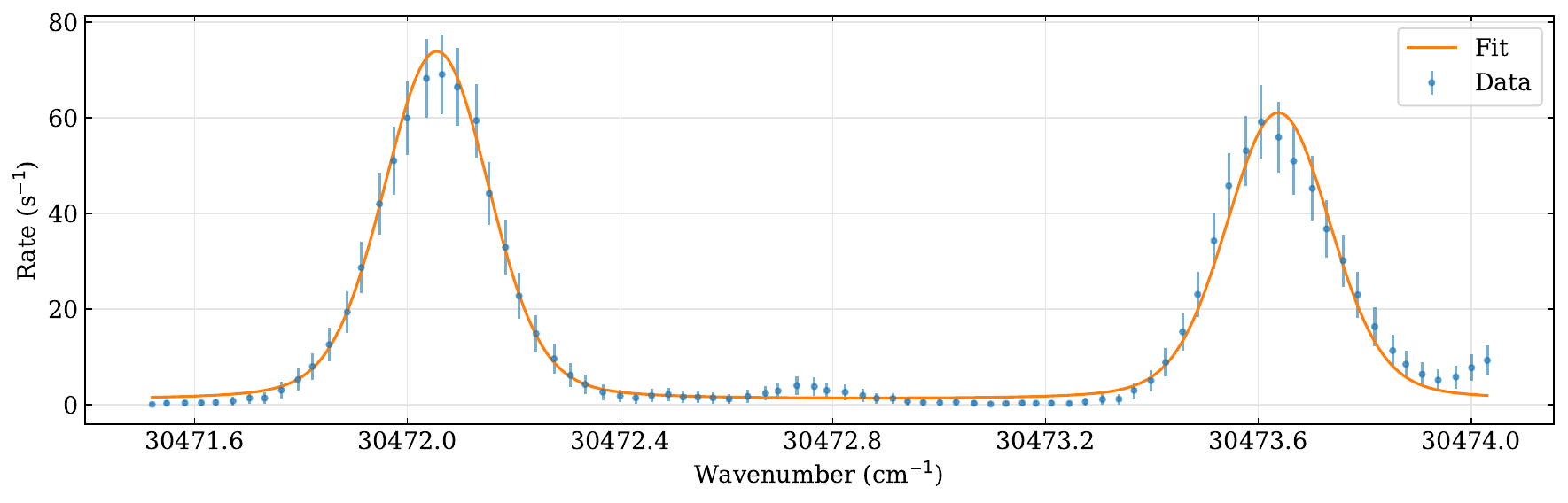}
\caption{In-source RIS hyperfine spectrum of $^{101}$Ag, fitted with the SATLAS2 package \cite{satlas2}. The two main peaks reflect the ground-state hyperfine splitting, with a small peak visible which may be the (1/2$^{-}$) isomer. See the text for details.} 
\label{fig:hfs}
\end{figure}

The performance and high sensitivity of this methodology have been successfully demonstrated using the highly challenging case of exotic, proton-rich silver isotopes approaching the $N=Z$ line. Here, a hyperfine spectrum is presented for the less-exotic isotope $^{101}$Ag, a nucleus with a ground state spin and parity $I^{\pi}=9/2^+$ (Fig.~\ref{fig:hfs}). Only a two-peak structure is resolved out of the six total hyperfine components due to a combination of laser linewidth and Doppler broadening within the hot cavity. However, this is sufficient for information to be extracted regarding fundamental properties of the nuclear structure: the hyperfine coupling constant ($A$), and thus the magnetic dipole moment ($\mu$), and the change in mean-squared charge radius ($\delta \langle r^2 \rangle$) with respect to a stable reference isotope. The Voigt profile fit indicated in Fig.~\ref{fig:hfs} results in a FWHM of $\approx$ 6.8 GHz, and a lower-state hyperfine constant A$_l$ = 9.79(9) GHz, agreeing within 1.4$\sigma$ with the value reported by Ferrer et al. (2014) of 9.64(6) GHz \cite{ferrer2014}. The fit does not take into account the possible isomer ($1/2^-$) seen between the main components of the ground state. Further analysis of in-source RIS hyperfine spectra of this and other isotopes, as well as implications for nuclear structure across the silver chain are ongoing.

\section{Summary and outlook}

In-source RIS in a hot-cavity catcher, combined with MR-ToF-based isobaric purification and detection, has been implemented at the IGISOL facility. In a first experiment, this method has been applied to proton-rich silver isotopes approaching the $N=Z$ line, extending the reach of previous of Penning trap-assisted measurements \cite{reponen2021}. The shorter purification timescale and higher efficiency of the MR-ToF-MS is particularly advantageous for the shortest-lived, most exotic cases. 

The proton-rich region of isotopic chains near $^{100}\text{Sn}$, the heaviest self-conjugate doubly magic nucleus, are of special interest, including the elements of tin, indium, cadmium, silver and palladium \cite{ge2024,karthein2024,sun2021}. This region remains a robust testing ground for nuclear structure theory, as it offers access to signatures of proton-neutron pairing along the $N=Z$ line, many isomeric states and it hosts the astrophysical rapid-proton-capture (rp-) process path \cite{wang2024,faestermann2013,parikh2013}. Further developments of the laser system and hot-cavity source aim to extend the reach of this technique, providing the basis for refining modern nuclear structure theories including DFT and $\it{ab\ initio}$ models across the $N=50$ shell closure.

\section{Acknowledgments}
{\normalsize
The Research Council of Finland (RCF) funded this work through the FIRI roadmap research infrastructures and  M.R.'s and M.K.'s Projects No. 368322 and 368323 (NUANZ). M.K. also acknowledges Centre of Excellence in Neutron-Star Physics project No. 374066. The European Union's Horizon 2020 research and innovation program grant agreement No. 861198–LISA–H2020-MSCA-ITN-2019 is acknowledged. A.K. acknowledges the funding from the RCF under grants No. 354968 (HENNA), No. 369714 (NUCLEI), and No. 374065 (Centre of Excellence in Neutron-Star Physics). Z.G. acknowledges RCF grant No. 354589. M.D. acknowledges the FWO PhD Fellowship Fundamental Research No. 11P6V24N.
}

\end{document}